\documentclass{aa}
\usepackage{graphicx}
\usepackage{txfonts}
\usepackage{natbib}
\usepackage{xcolor}

\usepackage{hyperref}

\hypersetup{
    colorlinks=true,        
    linkcolor=blue,        
    citecolor=blue,        
    urlcolor=blue
}

\begin{document} 

\title{Apophis source population and Earth encounter frequency of Apophis-like bodies}

\author{
M.~Bro\v{z}\inst{1}\and
R.P.~Binzel\inst{2}\and
P.~Vernazza\inst{3}\and
M.~Marsset\inst{4}\and
O.~Chrenko\inst{1}\and
J.~\v{D}urech\inst{1}\and
D.~Herald\inst{5}
}

\institute{
Charles University, Faculty of Mathematics and Physics, Astronomical Institute, V Holešovičkách 2, CZ-18000 Praha, Czech Republic
\and
Department of Earth, Atmospheric and Planetary Sciences, MIT, 77 Massachusetts Avenue, Cambridge, MA 02139, USA
\and
Aix-Marseille University, CNRS, CNES, LAM, Institut Origines, Marseille, France
\and
European Southern Observatory (ESO), Karl-Schwarzschild-Strasse 2, 85748 Garching bei München, Germany
\and
Trans-Tasman Occultation Alliance (TTOA), Wellington, PO Box 3181, New Zealand
}

\date{Received \today}

\abstract
{
Asteroid (99942) Apophis will safely pass the Earth in April 2029.
This extraordinary event will be observable by naked eye
from Africa and western Europe.
}
{
We provide context for Apophis' 2029 Earth passage by analyzing
possible source populations for bodies in Apophis-like orbits,
in particular, the Flora family,
which has a mineralogical composition
corresponding to LL chondrite meteorites,
similar to measurements of Apophis itself.
}
{
We estimated the specific encounter probability
of the present-day orbit of Apophis
by classical \"Opik and N-body methods.
We then performed orbital simulations of the Flora family,
initializing bodies between 2.1 and 2.3\,au,
and tracked their evolution into the near-Earth object (NEO) space
to compute their mean and peak encounter probabilities with Earth.
}
{
Out of an estimated population of ${\sim}3380$ NEOs larger or equal than Apophis
(${\ge}420\,{\rm m}$),
$610\pm 140$ are LL-like NEOs from Flora.
Their mean encounter probability is
$p = 86\times 10^{-18}\,{\rm km}^{-2}\,{\rm y}^{-1}$,
corresponding to once per 13000\,y frequency of encounters
closer than 38000\,km.
However, this does not apply to Apophis alone,
for which the specific encounter probability is higher
$p' = 1603\times 10^{-18}\,{\rm km}^{-2}\,{\rm y}^{-1}$,
but the frequency is lower,
only once per 430000\,y,
when we consider it as a single object.
Our simulation of the Flora family over $\sim$1 billion years
indicates that Apophis-like bodies from Flora have orbits
that are particularly persistent in near-Earth space.
The temporal distribution of encounter probabilities
exhibits peaks
(up to ${>}10^4$ in the same units)
and the specific value for Apophis is not unusual
(occurring ${\sim}70\%$ of time).
In other words,
there is always at least one Apophis-like body among NEOs.
We find that such persistence also creates favorable opportunities
for temporary capture as Earth coorbitals,
Apophis-like bodies are ultimately removed from the inner solar system
by approaching the Sun or by impact into one of the terrestrial planets,
where the relative split between these outcomes is
$(45\pm 2)\,\%$ and $(50\pm 2)\,\%$.
While our current knowledge of Apophis' orbit guarantees no threat from Apophis
in the next few centuries,
we cannot predict any specific outcome for Apophis
in the coming thousands or millions of years.
Evaluating this statistically over the long term,
we find that objects in Apophis-like orbits
have a $(19\pm 2)\,\%$ chance of Earth impact over their lifetime of ${\sim}30\,{\rm My}$.
}
{
Apophis appears to be a `prototypical' example
for the population of hundred-metre bodies
intersecting Earth's orbit
and impacting the Earth,
making it a particularly worthwhile target
for investigations advancing our knowledge
for planetary defence.
}


\maketitle


\section{Introduction}

The asteroid (99942) Apophis
is one of the near-Earth objects (NEOs)
that regularly undergoes close encounters with the Earth.
In this specific case, however, one of the encounters
will be particularly deep, on Friday, the 13th of April 2029
\citep{Giorgini_2008Icar..193....1G},
which explains its status of being a high-profile target for ongoing observational campaigns
\citep{Farnocchia_2013Icar..224..192F,Vokrouhlicky_2015Icar..252..277V,Souchay_2018A&A...617A..74S,Brozovic_2018Icar..300..115B,Binzel_2021BAAS...53d.045B,Reddy_2022PSJ.....3..123R}
and also for upcoming {\em in-situ\/} measurements
\citep{Polit_2024cosp...45..242P,Martino_2024cosp...45..241M}.

Given the importance of Apophis for planetary science and for planetary defence,
it is important to provide the corresponding context,
in terms of the overall NEO population.
Is Apophis exceptional?
Is the 2029 Apophis close encounter truly rare?
From where did Apophis most-likely originate?
Are there any other bodies among NEOs,
which can become ---on a long time scale --- Apophis-like?
These are the questions we address in this contribution.

The current, osculating orbit of Apophis
($a = 0.9227\,{\rm au}$, $e = 0.1914$, $i = 3.339^\circ$)
corresponds to a deep Earth-crosser,
which is detached from the belt between Mars and Jupiter.
Integrating its orbit backward in time and finding its origin within the belt
is impossible for multiple reasons
(deterministic chaos,
diffusion,
dissipative forces,
collisions,
$S = k.\log W$,
\dots).
Nevertheless,
most of the currently discovered NEOs are believed to originate from the Flora family,
located at the inner edge of the belt
\citep{Vernazza_2008Natur.454..858V,Marsset_2024Natur.634..561M,Broz_2024A&A...689A.183B,Broz_2024Natur.634..566B,Lagain_2025}.
This link is consistent with global modeling of transport from the belt
and with mineralogical classification of families and NEOs.
Specifically,
their mineralogy corresponds to LL chondrite meteorites.
Spectral observations of Apophis
also correspond to LL-like mineralogy
\citep{Binzel_2009Icar..200..480B,Reddy_2018AJ....155..140R},
so linking Apophis to Flora
seems to be consistent compositionally as well as dynamically.


\section{Results}

\subsection{Probability of coming from a source}\label{sec:source}

As a first step, we quantified probabilities of Apophis coming from
each of several relevant sources in the belt.
We used the METEOMOD/NEOMOD model%
\footnote{\url{https://sirrah.troja.mff.cuni.cz/~mira/meteomod/}}
from \cite{Broz_2024A&A...689A.183B,Lagain_2025},
which is based on an extensive set of 56 asteroid families.
For sub-km NEOs, only families are relevant,
because the background population is depleted
due to the low strength of sub-km bodies
\citep{Benz_1999Icar..142....5B,Bottke_2005Icar..175..111B,Bottke_2020AJ....160...14B}
and collisional cascade,
as confirmed by recent James Webb Space Telescope observations
\citep{Burdanov_2025Natur.638...74B}.
For each family, its synthetic size-frequency distribution (SFD),
$N_{\rm mb}({>}D)$,
is constrained by the observed SFD
and extrapolated to smaller sizes below the observational limit.
This extrapolation is not simplified (power-law),
but is based on a dedicated collisional model.
Then, the corresponding NEO population is estimated as
\begin{equation}
N_{\rm neo}({>}D) = {\tau_{\rm neo}\over\tau_{\rm mb}} N_{\rm mb}({>}D)\,,\label{eq:N_neo}
\end{equation}
where the two dynamical time scales
correspond to the mean lifetimes of synthetic orbits
in the belt and in the NEO space, respectively.
Our removal criteria were
$q < R_\odot$, $r > 100\,{\rm au}$.
(See more details in \citealt{Broz_2024A&A...689A.183B}.)

In order to estimate the source probability starting with any specific NEO orbit,
it turns out that the semi-major axis~$a$ and inclination~$i$
are better indicators than the orbital eccentricity~$e$,
because $e$ undergoes the greatest alteration.
($e$'s value must be substantially increased as a prerequisite to entering near-Earth space.)
Hence, binned distributions $M_j(a, i)$ of synthetic NEO orbits,
originating from individual families,
were used to compute the probability of coming from a source~$j$ as
\begin{equation}
p_j = N_{{\rm neo},j}({>}D) M_j(a, i)\,,
\end{equation}
and the sum $\sum_j p_j$ was re-normalized,
assuming there is no other, `mysterious' source.

If we use all families,
regardless of their mineralogical composition,
and assume $D = 1\,{\rm km}$ for simplicity,
we obtain the highest probability for Flora (${\sim}0.52$), followed by
Polana (0.29),
Nysa (0.09),
Juno (0.04), and
Vesta (0.04).
Generally, these are the families contributing most to the kilometre-size NEO population
and, simultaneously, producing objects on Apophis-like orbits.
We formally define Apophis-like bodies as NEOs having
perihelion $q < 1.016\,{\rm au}$,
aphelion $Q > 0.983\,{\rm au}$,
$a < 1.523\,{\rm au}$,
experiencing at least 5 close encounters (${<}0.01\,{\rm au}$)
with Earth per $10^4$ years, and
dynamical lifetime $\tau_{\rm neo} > 10\,{\rm My}$
(Fig.~\ref{apophislike}).

\begin{figure}
\centering
\begin{tabular}{c}
Flora (LL) \\
\includegraphics[width=9cm]{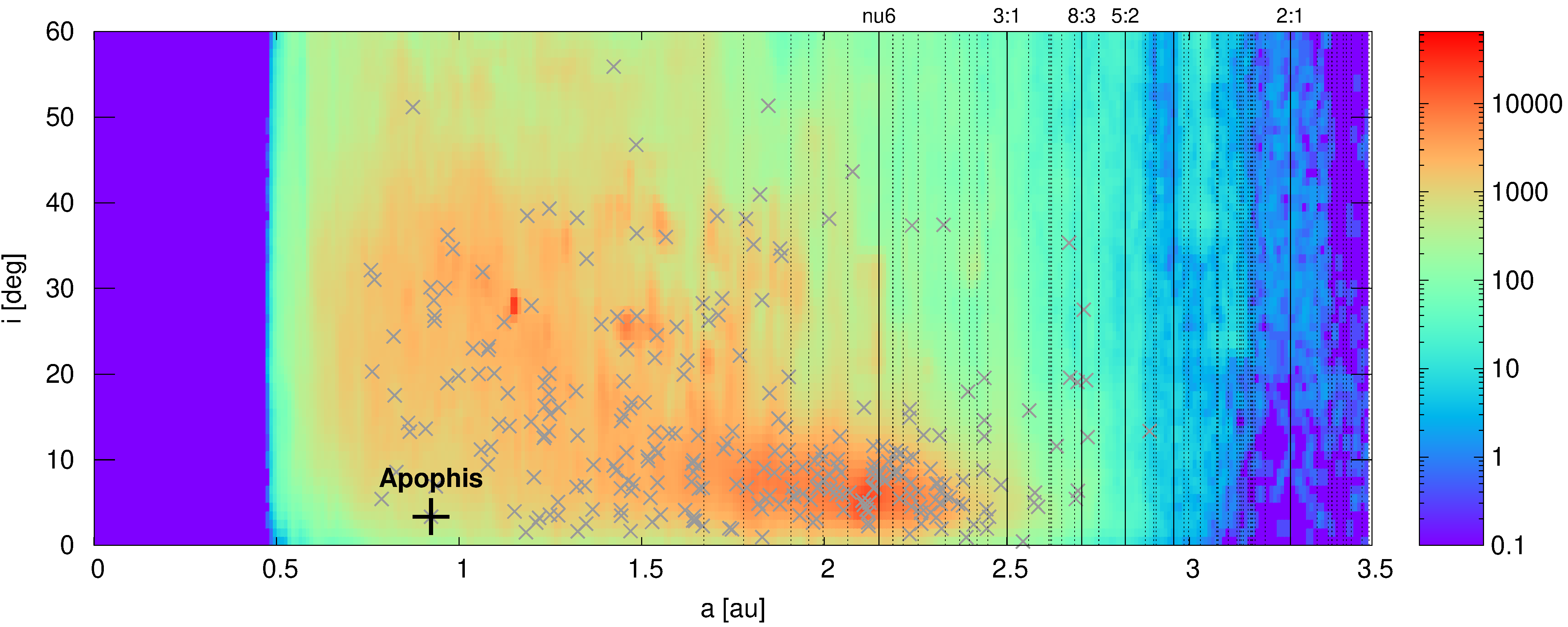} \\
Eunomia (LL) \\
\includegraphics[width=9cm]{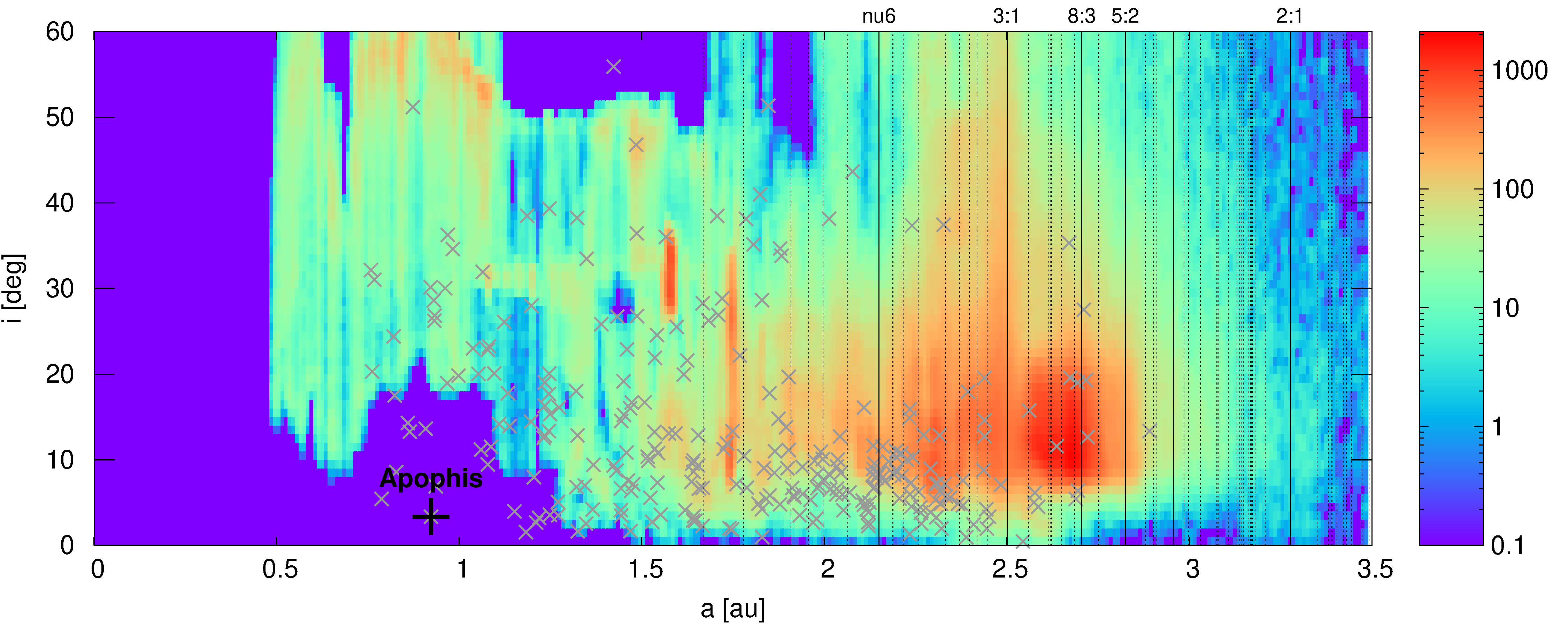} \\
Nysa (LL) \\
\includegraphics[width=9cm]{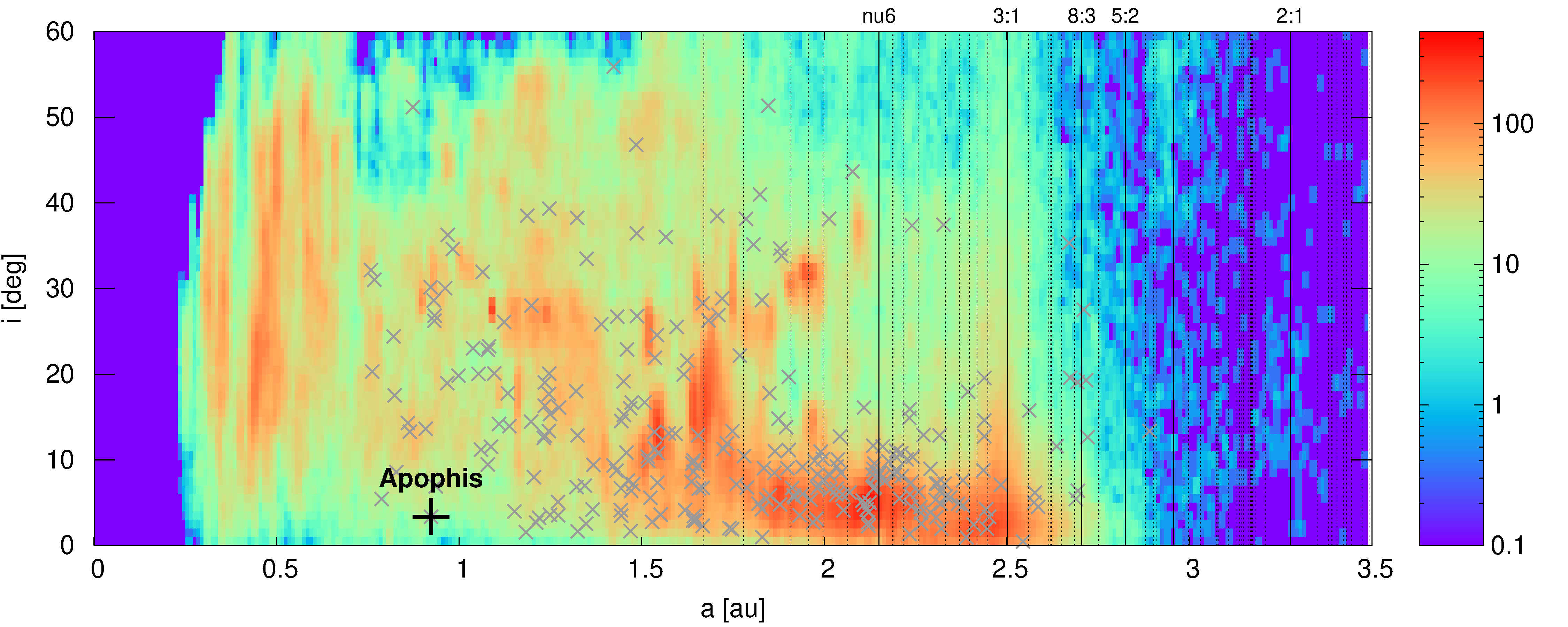} \\
\end{tabular}
\caption{
Binned distributions $M_j$ of semimajor axes $a$ and inclinations $i$
of synthetic NEO orbits,
originating from individual asteroid families.
Colours correspond to the number of test particles in bins;
although the quantity $M_j(a, i)$ was eventually normalized.
The bin sizes were $0.01\,{\rm au}$ and $1^\circ$, respectively.
Apophis orbit is marked with a black cross,
other LL-like NEOs from
\cite{Binzel_2019Icar..324...41B,Marsset_2022AJ....163..165M}
with gray crosses.
Of the three, the highest density for producing Apophis-like objects is from the Flora family.}
\label{maps}
\end{figure}

Mineralogical composition provides an additional constraint
on the source population of Apophis.
Since it is spectroscopically analogous to LL chondrite meteorites,
we restricted our search to the subset of LL-like source regions identified by
\cite{Vernazza_2008Natur.454..858V,Marsset_2024Natur.634..561M}.
These include only three major asteroid families,
Flora,
Eunomia,
and a part of Nysa,
since another part of Nysa is either too dark (M-type) or too bright (E-type).
We also considered the Juno family,
whose members are mostly identified compositionally as L/LL \citep{Marsset_2024Natur.634..561M}.
The resulting probabilities are summarized in Tab.~\ref{tab:meteomod}.
Only Flora, Nysa and Juno produced some Apophis-like orbits,
because Eunomia is located a bit too far,
beyond the 3:1 mean-motion resonance with Jupiter
(see Figs.~\ref{maps}, \ref{juno}).
Furthermore, Flora (with ${\sim}0.86$ probability)
dominates other sources,
not only because Flora family members are so numerous,
but it is also best-positioned source
due to the proximity of the $\nu_6$ secular resonance
\citep{Vokrouhlicky_2017AJ....153..172V}
and the relatively strong influence of the Yarkovsky effect at the inner edge of the asteroid belt at 2.1\,au
\citep{Vokrouhlicky_1999AJ....118.3049V}.

\begin{table*}
\caption{
Source probabilities from five asteroid families (including three dominated by LL compositions) for delivering objects into Apophis-like orbits.
}
\label{tab:meteomod}
\small
\centering
\begin{tabular}{lrrrrrrl}
             &                  & 420-m           & 420-m        & 420-m         &            &       &       \\
\vrule width 0pt depth 4pt
source       & $\tau_{\rm neo}$ & $\tau_{\rm mb}$ & $N_{\rm mb}$ & $N_{\rm neo}$ & $M_j(a,i)$ & $p_j$ & notes \\
-- & My & My & $10^3$ & 1 & 1 & 1 & -- \\
\hline
\vrule width 0pt height 9pt
Vesta (HED)     & 4.39  &  856 & 27.0 &  138.5     & $1.8\times 10^{-5}$  & --    & calibration \\ 
Juno (L/LL)     & 2.55  &  259 & 15.0 &  147.7     & $2.4\times 10^{-5}$  & 0.089 & \\ 
Flora (LL)      & 9.95  &  340 & 21.0 &  614.6     & $4.9\times 10^{-5}$  & 0.761 & \\ 
Eunomia (LL)    & 4.48  & 1539 & 17.5 &   50.9     & $0.0\times 10^{-5}$  & 0     & \\ 
Nysa (LL)       & 4.04  &  394 & 17.4 &  178.4     & $3.3\times 10^{-5}$  & 0.148 & including E, M \\ 
\hline
\vrule width 0pt height 9pt
all NEOs        &       &      &      & $\sim$3380 &                      &       & \cite{Nesvorny_2024Icar..41716110N} \\
\end{tabular}
\end{table*}



\subsection{Mean encounter frequency for Apophis}\label{sec:mean}

In order to estimate the mean encounter frequency for Apophis,
one needs to know its size,
because the population is size-dependent.
Unfortunately, there is a tangible disagreement between the volume-equivalent diameters
340\,m from radar \citep{Brozovic_2018Icar..300..115B}
vs. 380\,m from infrared observations \citep{Muller_2014A&A...566A..22M}
vs. 420\,m from occultations
(Appendix~\ref{sec:durech}).
Hereinafter, we prefer the latter, since occultations
are less prone to calibration or systematic uncertainties, provided
the respective occultation event was not too short,
cadence of measurements not too long
or that diffraction patterns were taken into consideration
(e.g, \citealt{Broz_2021}).

The cumulative number of bodies $N_{\rm mb}$ in the Flora family,
which are larger than 420\,m,
was extrapolated from the observed size distribution
using a collisional model of \cite{Broz_2024A&A...689A.183B,Lagain_2025}.
The equilibrium NEO population originating from Flora was estimated
again from Eq.~(\ref{eq:N_neo}).
For
$N_{\rm mb} = 21000\pm 2000$,
$\tau_{\rm neo} = (9.95\pm 0.5)\,{\rm My}$, and
$\tau_{\rm mb} = (340\pm 30)\,{\rm My}$,
we obtained
$N_{\rm neo} = 610\pm 140$.
The uncertainty of this estimate is mostly due to
stochasticity of the collisional model at sub-km sizes.





We estimated the corresponding collisional probabilities of these bodies and
the frequency of encounters with the Earth.
From our N-body simulations of transport from Flora
\citep{Broz_2024A&A...689A.183B,Lagain_2025},
the intrinsic, scaled, collisional probability of NEOs is
\begin{equation}
p = (86.212\pm 3) \times 10^{-18}\,{\rm km}^{-2}\,{\rm y}^{-1}\,,\label{eq:p}
\end{equation}
hence the flux
\begin{equation}
\Phi = p N_{\rm neo} = (5.3\pm 1.4) \times 10^{-14}\,{\rm km}^{-2}\,{\rm y}^{-1}
\end{equation}
and the number of encounters
\begin{equation}
N_{\rm enc} = \Phi \Delta t R^2\,,
\end{equation}
where we used $N_{\rm neo}$ as the number of pairs
(since Earth is just one).
If we assume the geocentric distance $R = 38000\,{\rm km}$,
the interval is approximately $\Delta t = (13000\pm 3500)\,{\rm y}$,
in order to get $N_{\rm enc} = 1$.
This is a longer mean interval than the 7500\,y value from
\citet{Farnocchia_2021RNAAS...5..257F,Nesvorny_2024Icar..41716110N},
because here we assumed a specific sub-population of LL-like NEOs from Flora.

So, does this make the 2029 Apophis encounter an even greater rarity
than previously calculated?
Well, no, because the value above is rather a characteristic of the NEO population,
not of Apophis alone.
Before answering the question, one needs to know three things:
(i)~What is the specific collisional probability for Apophis?
(ii)~What is the long-term evolution of Apophis' orbit?
(iii)~What are the encounter frequencies for Apophis-like bodies?


\subsection{Specific encounter frequency for Apophis}\label{sec:specific}

If the current, osculating orbit of Apophis is used
to compute its collisional probability
with the \"Opik theory
\citep{Opik_1951PRIA...54..165O,Bottke_1993GeoRL..20..879B},
one obtains
\begin{equation}
p' = 1603 \times 10^{-18}\,{\rm km}^{-2}\,{\rm y}^{-1}\,.\label{eq:p_}
\end{equation}
This is almost 20 times more than $p$ above
(Eq.~(\ref{eq:p})),
a confirmation that Apophis orbit is exceptional,
at the current epoch,
compared to other, ordinary NEOs.
If we focus on Apophis alone ($N_{\rm neo} = 1$),
its specific, single-body encounter frequency would be as long as
once per $430000\,{\rm y}$.

Note the \"Opik theory assumes a uniform precession of angles
($\omega$, $\Omega$),
which is a good approximation of short-term orbital evolution,
provided the orbits of colliding bodies are uncorrelated.
This is most likely the case for old families and most NEOs
(but not for young families; \citealt{Vokrouhlicky_2021A&A...654A..75V}).
Of course, for ephemerides and deterministic predictions of encounters
(2029, 2036, 2068, \dots),
this is not a suitable theory,
but here we are interested in statistics of encounters.


\subsection{Long-term evolution of Apophis orbit}\label{sec:longterm}

We also computed the long-term evolution of clones having Apophis-like orbits,
using a standard N-body model
\citep{Levison_1994Icar..108...18L,Broz_2011MNRAS.414.2716B}.
We employed the symplectic integrator RMVS3,
which handles close encounters with planets
by subdividing the time step.
We included 
gravitational perturbations by planets from Mercury to Neptune, Ceres, Vesta,
the Yarkovsky effect,
the YORP effect;
assuming simplified, principal-axis rotation,
with the period $P = 30.56\,{\rm h}$,
but not tumbling \citep{Pravec_2014Icar..233...48P,Lee_2022A&A...661L...3L},
which is an acceptable approximation \citep{Vokrouhlicky_2015Icar..252..277V}.
The respective thermal parameters correspond to S-type bodies
covered by regolith,
the bulk density $\rho = 2000\,{\rm kg}\,{\rm m}^{-3}$,
the conductivity $K = 0.07\,{\rm W}\,{\rm m}^{-1}\,{\rm K}^{-1}$, and
the capacity $C = 680\,{\rm J}\,{\rm kg}^{-1}\,{\rm K}^{-1}$,
the Bond albedo $A = 0.14$,
the emissivity $\varepsilon = 0.86$.
The diurnal drifts $\dot a$ then span a range
$\pm0.0015\,{\rm au}\,{\rm My}^{-1}$
for various obliquities.
For comparison,
the actual astrometric measurement of the Yarkovsky drift for Apophis
is within this range, $\dot a = (-0.00133\pm 0.00001)\,{\rm au}\,{\rm My}^{-1}$
\citep{Perez_2022ComEE...3...10P,Farnocchia_2022LPICo2681.2007F}.

\begin{figure}
\centering
\begin{tabular}{l}
\includegraphics[width=9cm]{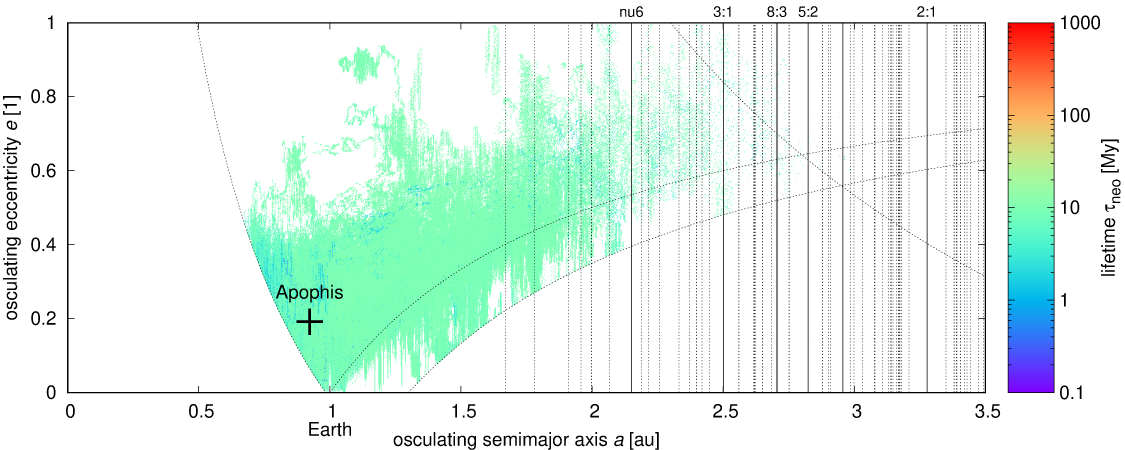} \\
\includegraphics[width=8.1cm]{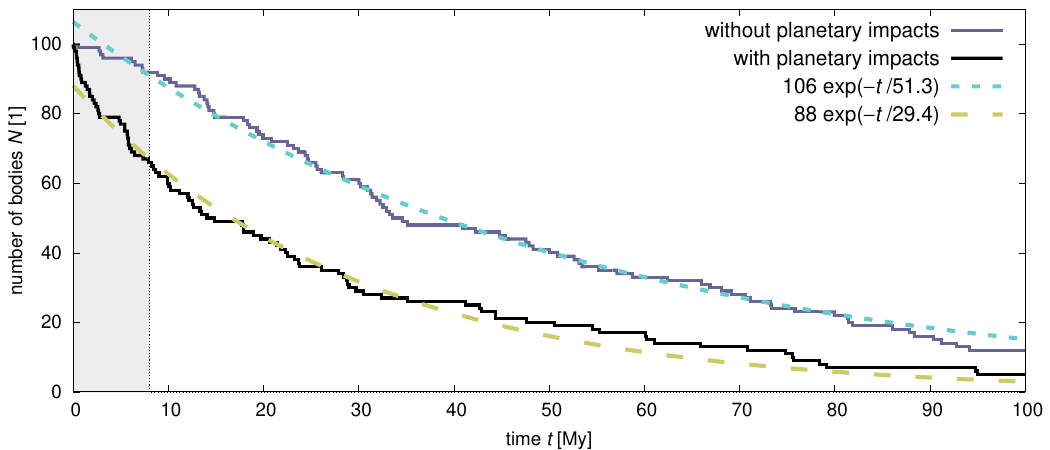} \\
\end{tabular}
\caption{
Top: Osculating semimajor axis $a$ versus eccentricity $e$
showing long-term evolution of Apophis-like orbits, after 10\,My.
The curves correspond to the boundaries of the NEO space
($q < 1.3\,{\rm au}$, $Q > 0.983\,{\rm au}$)
and of the adjacent Mars-crossing population.
The dotted vertical lines correspond to locations
of various mean-motion resonances,
including the $\nu_6$ secular resonance.
Majority of the 100 clones of Apophis
remain in the NEO (or Mars-crossing) space;
the respective orbit is relatively long-lived.
Bottom: The number of Apophis clones $N$ versus time $t$.
The corresponding e-folding time scale is ${\sim}30$\,My,
or $50$\,My respectively,
depending on whether or not planetary impacts were used
as a criterion for removal of bodies.
}
\label{apophis-2_nea2}
\end{figure}

We used 100 clones of Apophis,
with various obliquities.
The time step was 0.9\,d,
the output time step $\delta t = 10^3\,{\rm y}$, and
the overall time span $\Delta t \ge 10\,{\rm My}$,
in order to obtain a statistical distribution of up to $10^6$ sets of osculating orbital elements,
shown in Fig.~\ref{apophis-2_nea2}.
We computed the collisional probability for this sample and obtained
\begin{equation}
p'' = 452\times 10^{-18}\,{\rm km}^{-2}\,{\rm y}^{-1}\,.\label{eq:p__}
\end{equation}
This is still almost an order of magnitude more than $p$ above
(Eq.~(\ref{eq:p}));
a confirmation that Apophis' orbit is exceptional,
even over its long-term evolution.
We find that objects in Apophis-like orbits spend more time having
on average a relatively higher collisional probability,
when compared to other NEOs.
We note the value is temporally dependent,
as $p''(t)$ tends to decrease,
as an Apophis-like orbit evolves across the NEO space.

Moreover,
the dynamical lifetime of some orbits is up to $100\,{\rm My}$
(Fig.~\ref{apophis-2_nea2}, bottom).
The clones were eliminated slowly,
with e-folding time scale ${\sim}30\,{\rm My}$,
often by being too close to the Sun or by planetary impacts.
This is again above the average ${\sim}8\,{\rm My}$
\citep{Granvik_2016Natur.530..303G,Granvik_2018Icar..312..181G}.
The reason is that at moderate inclinations of at least few $^\circ$,
and not crossing any other terrestrial planets
($q = 0.7460\,{\rm au}$, $Q = 1.0993\,{\rm au}$),
collisional probabilities decrease
and an NEO orbit can survive for a long time.
In other words, we find that on average, objects in Apophis-like orbits
persist longer than most other objects in near-Earth space.

\subsection{Peak encounter frequency for Apophis-like bodies}\label{sec:peak}

As far as Apophis-like bodies are concerned
(cf.~Fig.~\ref{apophislike}),
these clearly originate from the Flora family.
We thus took the synthetic Flora family from
\cite{Broz_2024A&A...689A.183B,Lagain_2025}
and analyzed the orbital distribution of the corresponding synthetic NEOs.
Since the age of the Flora family is
$(1.3\pm 0.1)\,{\rm Gy}$,
in the past,
it supplied substantially more objects to the NEO population than today
\citep{Vokrouhlicky_2017AJ....153..172V,Lagain_2025}.
In our simulation, we started with an arbitrary, representative sample
of $\sim$1900 test particles
and their number decreased slowly, exponentially, in the course of time.

\begin{figure}
\centering
\includegraphics[width=9cm]{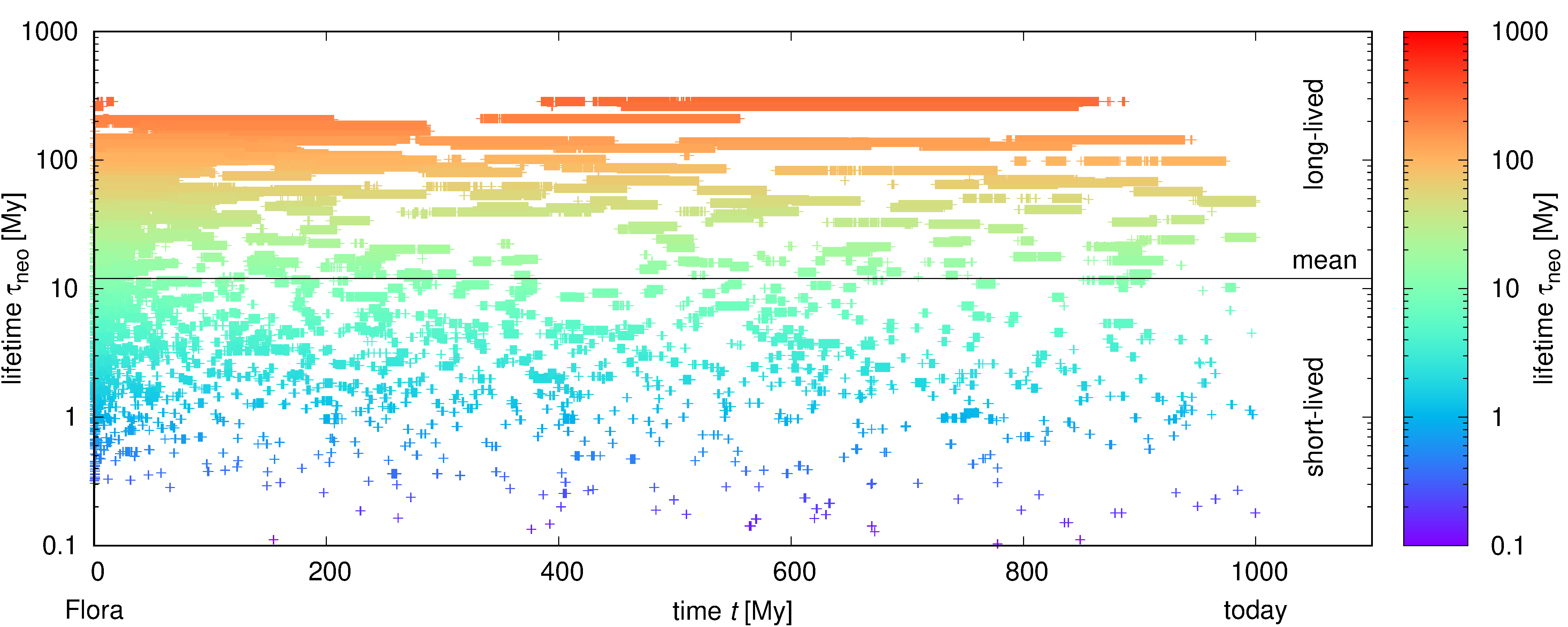} 
\caption{
Lifetimes of LL-like NEOs originating from the Flora family
according to our simulation.
The $x$-axis is simulation time
(spanning from 0, i.e., Flora disruption, to 1\,Gy, the end of simulation),
the $y$-axis is lifetime.
One can see both short- (\textcolor{blue}{blue})
and long-lived orbits (\textcolor{orange}{orange});
their `mix' determines the mean life time $\tau_{\rm neo}$.
}
\label{flora-1_300MS_nea4}
\end{figure}

\begin{figure}
\centering
\begin{tabular}{ll}
\includegraphics[width=9cm]{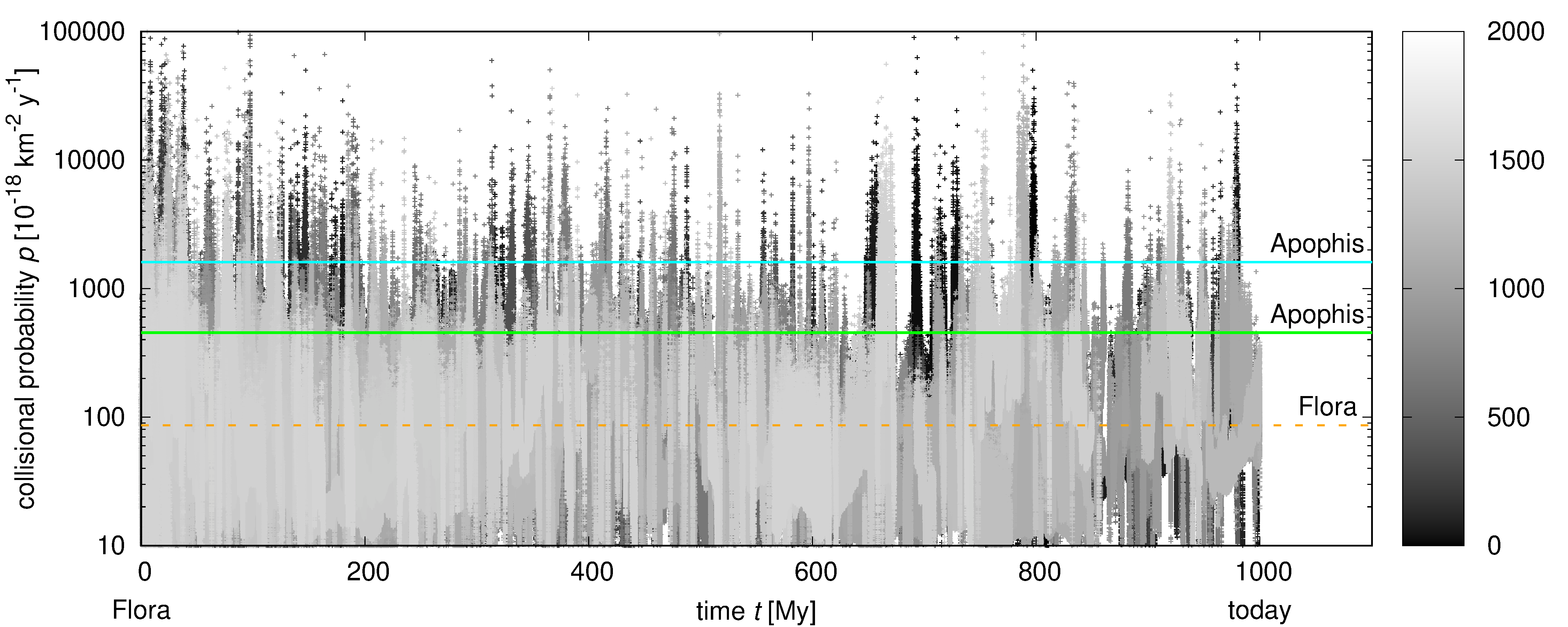} \\
\includegraphics[width=9cm]{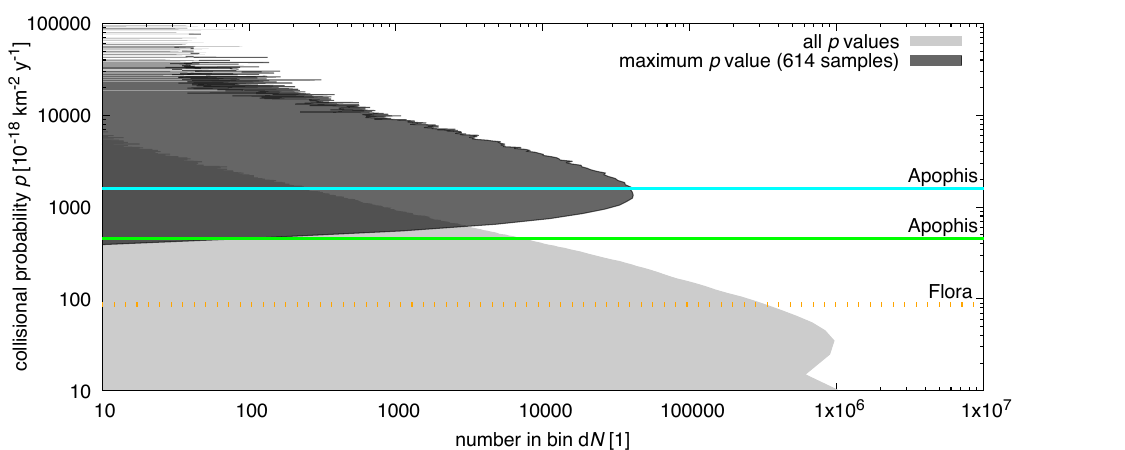} \\
\end{tabular}
\caption{
Top: Collisional probabilities of LL-like NEOs originating from the Flora family
and their temporal distribution.
Individual bodies (out of 1900) are distinguished by shades of gray.
Vertical range shows both small and large probabilities.
For comparison, horizontal lines indicate
the mean probability~$p$ of all NEOs (\textcolor{orange}{orange}),
short-term, initial probability $p'$ of Apophis (\textcolor{cyan}{cyan}),
long-term probability $p''$ of Apophis (\textcolor{green}{green}).
Bottom: The distribution of all $p$ values (\textcolor{gray}{gray})
and of maximum $p$ value (black).
}
\label{flora-1_300MS_proba_t}
\end{figure}

Out of these, one can distinguish short- and long-lived orbits
(their `mix' determines the mean life time $\tau_{\rm neo}$).
The actual life times of individual bodies are diverse,
ranging from ${<}0.1$ to hundreds of My
(see Fig.~\ref{flora-1_300MS_nea4}).
Of course, this is closely related to collisional probabilities,
which are also diverse,
from zero (if orbits do not intersect the Earth) up to ${>}10^4$,
in the $10^{-18}\,{\rm km}^{-2}\,{\rm y}^{-1}$ units
(see Fig.~\ref{flora-1_300MS_proba_t}).
Such peaks are common and there are intervals of time,
for which at least one Apophis-like body is statistically present among NEOs,
throughout the entire duration of our simulation.

Finally, we must take into account that the actual number of bodies
larger than $420\,{\rm m}$
is higher than in our simulation.
Since the peaks in probabilities are randomly distributed in time,
one can simply use a Monte-Carlo approach:
draw a sample of $N_{\rm neo} = 610$ from all $p$ values,
determine its maximum $p$ value, and
repeat to obtain the respective distribution.
According to Fig.~\ref{flora-1_300MS_proba_t} (bottom),
there is a ${\sim}70\%$ chance of having
at least one Apophis-like body
in the population of LL-like NEOs.


\subsection{The fate of (99942) Apophis}

More than two decades of orbital tracking,
including high precision radar position and velocity measurements,
conclusively show that Apophis poses absolutely no threat to Earth
in the coming few centuries \citep{Farnocchia_2022LPICo2681.2007F,Brozovic_2022LPICo2681.2023B,Chesley_2023LPICo2988.2031C,Dotson_2023LPICo2988.2035D}. Nevertheless, the ultimate fate of Apophis in a forecast of thousands or millions of years is unknown.

\begin{figure}
\begin{tabular}{ll}
\includegraphics[width=8cm]{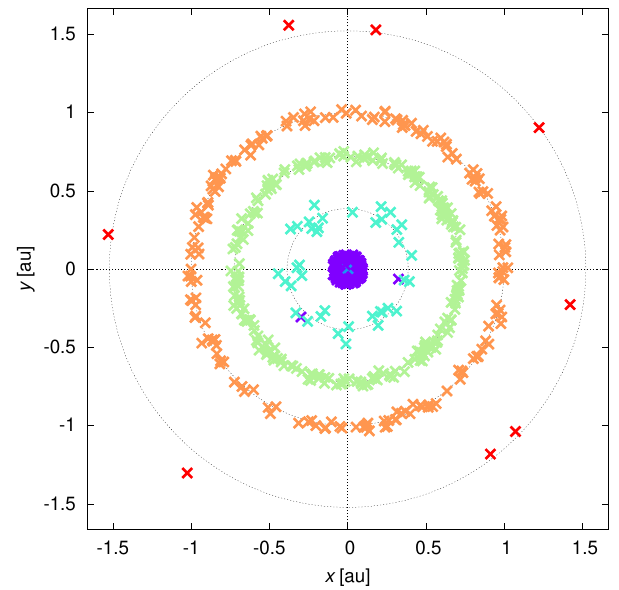} \\
\kern.1cm\includegraphics[width=7.8cm]{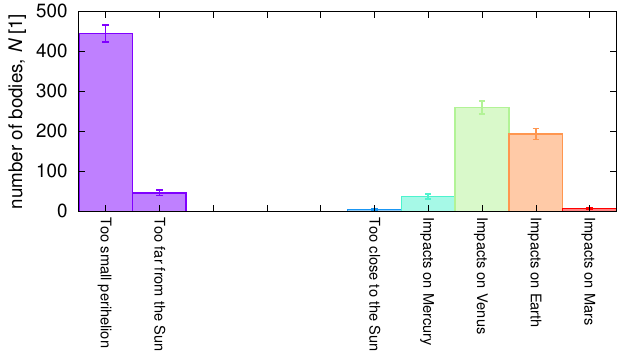} \\
\end{tabular}
\caption{
Top: Collisions of Apophis clones with the Sun, Mercury, Venus, Earth, or Mars.
Their positions in the ecliptic frame ($x$, $y$) are indicated (crosses),
alongside the mean orbital distances of terrestrial planets (dotted circles).
Bottom: Statistics of collisions.
Specifically, $(45\pm 2)\,\%$ of the clones were eliminated due to low perihelion
($q < R_\odot$),
while $(50\pm 2)\,\%$ due to planetary impacts.
The respective uncertainties (error bars) are Poissonian (${\propto}\sqrt{N}$).
}
\label{discard}
\end{figure}

To evaluate the statistical possibilities in a long-term Apophis forecast,
we used our long-term orbital simulations
(Secs.~\ref{sec:longterm}, \ref{sec:peak})
to understand the fate of bodies on Apophis-like orbits.
We evaluated the outcomes of Apophis clones
by considering their statistical removal from the inner solar system
through close intersections or collisions with
the Sun, Mercury, Venus, Earth, and Mars
(Fig.~\ref{discard}).
It turned out that over the typical 30 million year lifetime for an Apophis clone,
approximately $(45\pm 2)\,\%$ of the clones
were eliminated due to low perihelion
($q < R_\odot$),
and $(50\pm 2)\,\%$ due to planetary impacts.
Out of those, Earth and Venus were the most effective planets,
since they have the most sizable cross sections
and focussing factors.
Such increased probabilities of planetary impacts
were previously reported for some `special' NEOs, like
ejecta from the Moon-forming impact
\citep{Bottke_2014LPI....45.1611B},
space debris
\citep{Rein_2018Aeros...5...57R}, or
Eros
\citep{Michel_1998AJ....116.2023M}.
This is very different from other, ordinary NEOs,
which mostly collide with the Sun,
and only ${\sim}1\%$ of them impact the Earth
(Fig.~\ref{nea-3_420m_lclose_0.9d}).

Our findings thus reveal the particular importance of objects in Apophis-like orbits
for evaluating the non-negligible long-term impact hazard to Earth.
Most profoundly, they indicate the scientific study opportunity
presented by the 2029 close passage by Apophis
is particularly valuable toward advancing our knowledge
in the applied science of planetary defence.

\section{Discussion}

Another implication of our long-term evaluation of Apophis-like orbits is considering the frequency with which such objects may pass by Earth at a near-miss distance inside the Roche limit.

\subsection{Passages within the Roche limit}\label{roche}

We thus computed the frequency of passages within the Roche limit.
Specifically, for Apophis-like objects:
\begin{equation}
R_{\rm roche} = k R_\oplus \left({\rho\over\rho_\oplus}\right)^{1\over 3} \approx 17900\,{\rm km}\,,
\end{equation}
where we used
$\rho = 2000\,{\rm kg}\,{\rm m}^{-3}$
\citep{Vokrouhlicky_2015Icar..252..277V};
the coefficient $k \approx 2$ suitable for rubble-piles
is lower than for fluids
\citep{Leinhardt_2012MNRAS.424.1419L,Zhang_2020A&A...640A.102Z}.
Since we already know the frequency of
$R \ge 38000\,{\rm km}$
encounters (Sec.~\ref{sec:specific}),
the surface area ratio is
$(R_{\rm roche}/R)^2 = 0.22$,
the frequency of Roche-limit encounters is
once per 1.9\,My.
However, this is a relatively long time
and the long-term collisional probability $p''$
(Eq.~(\ref{eq:p__}))
is lower,
which shifts the estimate to
once per 6.9\,My.
Consequently, it is possible that Apophis has undergone in the past, or will undergo in the future, several Roche-limit passages
over the course of its lifetime
(${\sim}30\,{\rm My}$)
as a near-Earth object.
The non-zero number ($\sim$3 to 7) of Roche-limit passages
is important for interpretation of {\em in situ\/} observations.

Such passages make asteroid surfaces look more fresh, or Q-type
\citep{Binzel_2010Natur.463..331B,Nesvorny_2010Icar..209..510N}.
Surface refreshment may come from seismic shaking during the encounter,
or subsequent long-term re-equilibration
after finding itself in a tidally-torqued new spin state
\citep{Ballouz_2024PSJ.....5..251B}.
At the same time, however, these orbits often exhibit low perihelia,
which leads to thermal processing of regolith
\citep{Delbo_2014Natur.508..233D,Graves_2019Icar..322....1G}.
Apophis itself has a weathered, transitional, Sq-type spectrum
\citep{Binzel_2009Icar..200..480B},
between S and Q,
which would be an indication of relatively long (${>}1\,{\rm My}$) interval
from the last resurfacing event.


\subsection{Relation to temporary coorbitals}\label{sec:coorbitals}

\begin{figure}
\centering
\includegraphics[width=8cm]{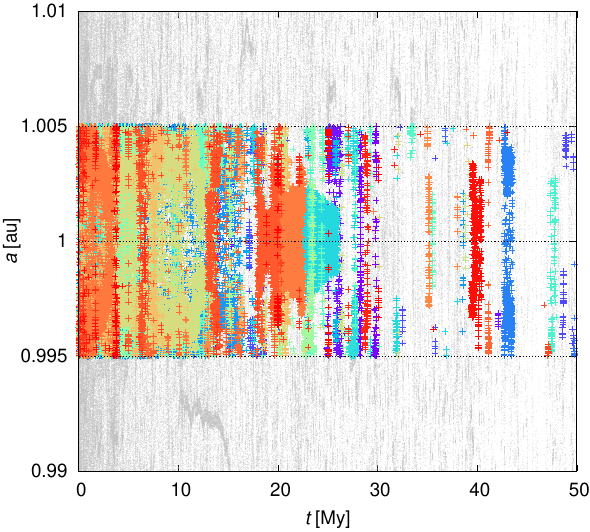}
\caption{
Osculating semimajor axis $a$ versus time $t$ for Apophis clones.
When bodies were located within the coorbital region of the Earth
($a \in (0.995; 1.005)\,{\rm au})$, $e < 0.2$),
they were plotted in colour to distinguish individual bodies,
otherwise in gray.
}
\label{coorbitals}
\end{figure}

In order to understand possible relation of Apophis-like bodies
to other NEO populations,
we also checked how often they become Earth coorbitals
\citep{Wiegert_2000Icar..145...33W,Christou_2000Icar..144....1C}.
According to the long-term evolution of Apophis clones
(Sec.~\ref{sec:longterm}),
415 out of 1000 bodies
were temporarily located in the coorbital region
($a \in (0.995; 1.005)\,{\rm au}$, $e \lesssim 0.2$).
Some captures were short
(${\le}10^3\,{\rm y}$, i.e., corresponding to our sampling),
but we also noticed ten long-term captures,
spanning one or more~My
(see Fig.~\ref{coorbitals}).
They orbited on horseshoe orbits,
which is the most common type
\citep{Kaplan_2020MNRAS.496.4420K}.
The percentage of time Apophis clones spent in this region
is of the order of $0.5\pm0.1$\,\%.
For comparison, all NEOs larger than 420\,m would spend
only 0.02\,\% as coorbitals,
because they, on average, do not orbit so close to the Earth.
An implication is that Apophis-like bodies
are the most likely source of Earth coorbitals.%

From where else Earth's coorbitals could come from?
It must be from within the NEO region,
because one cannot inject an object into the coorbital region
without being temporarily a NEO.
Moreover, the respective mechanism must provide
sufficient~$\Delta v$,
like close encounters with the Earth (or Venus);
pointing again to Apophis-like bodies.%
\footnote{
The lunar origin suggested for some coorbitals
\citep{Jiao_2024NatAs...8..819J}
is not feasible for kilometre-sized objects.
}

\begin{figure}
\centering
\includegraphics[width=9cm]{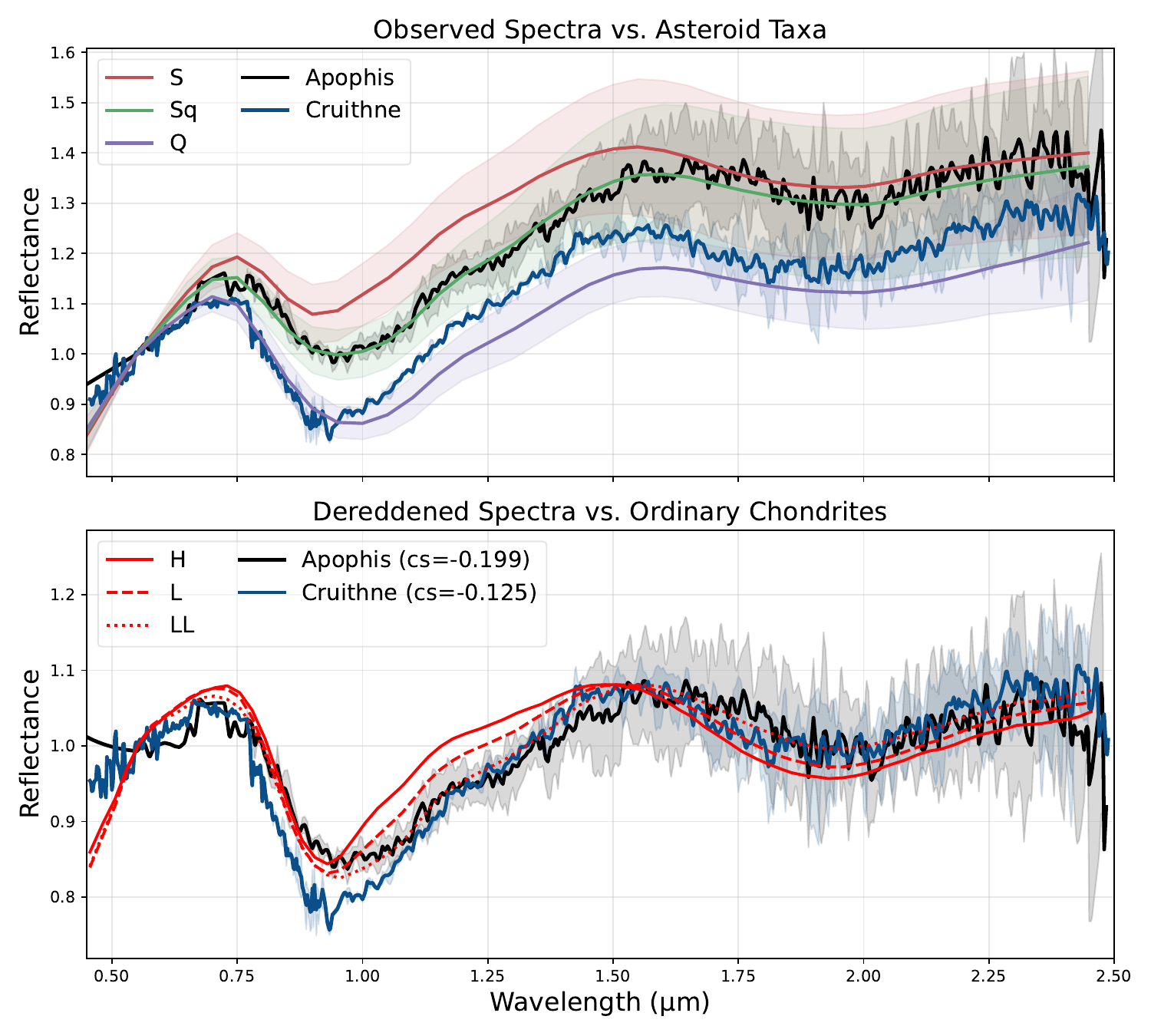}
\caption{
Top: Observed mean reflectance spectrum of Apophis and Cruithne
from the MITHNEOS survey
\citep{Binzel_2019Icar..324...41B,Marsset_2022AJ....163..165M}.
Their comparison to taxonomic classes (S, Sq, Q) shows that
Apophis is Sq-type and Cruithne Q-type.
Bottom: The same spectra after dereddening
\citep{Brunetto_2006Icar..184..327B};
note the applicability in the visible of this simplified,
$\exp(-C_{\rm S}/\lambda)$ model is limited.
Their comparison to ordinary chondrite classes (H, L, LL) shows that
both asteroids are LL-like.
}
\label{marsset}
\end{figure}

This is in agreement with spectral observations of
(3753) Cruithne
\citep{Angeli_2002A&A...391..757A,Deleon_2010A&A...517A..23D},
the only known kilometre-sized Earth coorbital
\citep{Usui_2011PASJ...63.1117U}.
Cruithne itself does not qualify as Apophis-like by our criteria,
as it currently experiences no close encounters with Earth,
despite its coorbital configuration.
Nevertheless, it is classified as Q-type,
suggesting a relatively recent (${<}1\,{\rm My}$)
resurfacing (Sec.~\ref{roche}).
We determined mineralogical composition of Cruithne
from its reflectance spectrum
\citep{Binzel_2019Icar..324...41B,Marsset_2022AJ....163..165M},
using the
\citet{Brunetto_2006Icar..184..327B}
and
\citet{Shkuratov_1999Icar..137..235S}
models for reddening and reflection, respectively
(Fig.~\ref{marsset}).
The ratio of olivine and orthopyroxene,
ol/(ol+opx) = 0.755,
corresponds to LL-like composition, 
fully compatible with the Flora family origin
(cf. \citealt{Marsset_2024Natur.634..561M}, fig.~1).

\section{Conclusions}

We reach a similar rarity of once per $\sim$4700~years
as \citet{Farnocchia_2021RNAAS...5..257F}
for the 2029 close passage by (99942) Apophis
(see Appendices~\ref{farnocchia} and~\ref{nbody}),
and we find that the Flora family is rather prolific
as the progenitor for such encounters by LL-like NEOs,
orbiting on relatively long-lived ($\sim$30\,My) orbits
in Earth's vicinity.
While Apophis-like bodies represent only ${\sim}3\%$
of the NEO population larger than 420\,m,
they are responsible for approximately half of all close Earth encounters.
Apophis, as a `prototypical' example, will end up
its journey
i)~by being removed by Earth or Venus, or slightly less likely,
ii)~by approaching too closely the Sun.
An implication is that these bodies
pass several times within the Roche limit
and because their orbits are so similar to Earth,
they are also related to temporary coorbitals.
Flora will nonetheless continue to contribute additional NEOs,
which will undergo deep, close encounters with planets.
This behaviour is clearly seen in our simulations
of the Flora family.

Ad communication.
If considering Apophis alone,
its encounter frequency with Earth is once per $\sim$430000\,y.
It is very, very rare, because it was computed
for a single body and
for a close encounter within the geocentric distance
$R = 38000\,{\rm km}$.
If considering LL-like NEOs,
originating from the Flora family,
their encounter frequency is once per $\sim$13000\,y.
Only if one considers all NEOs,
regardless of their mineralogy,
the frequency is once per $\sim$4700\,y.

\vskip\baselineskip

\begin{table}
\caption{
Estimates of populations $N_{\rm mb}({>}D)$, $N_{\rm neo}({>}D)$
and encounter frequencies $\Delta t$
for Apophis itself,
LL-like NEOs from Flora,
and all NEOs.
}
\label{tab2}
\centering
\small
\begin{tabular}{lll}
Apophis $D$             & {\bf 420\,m}                     & \textcolor{gray}{380\,m}             \\
\hline
\vrule width 0pt height 9pt
Apophis $\Delta t$      & $\mathbf{430000}${\bf\,y}        & \textcolor{gray}{430000\,y}          \\
Flora $N_{\rm mb}$      & $\mathbf{21000\pm 2000}$         & \textcolor{gray}{$23000\pm 2000$}    \\
Flora $N_{\rm neo}$     & $\mathbf{610\pm 140}$            & \textcolor{gray}{$670\pm 150$}       \\
Flora $\Delta t$        & $\mathbf{13000\pm 3500}${\bf\,y} & \textcolor{gray}{$12000\pm 3100$\,y} \\ 
NEOs $N_{\rm neo}$      & $\mathbf{3380\pm 300}$           & \textcolor{gray}{$3980\pm 350$}      \\
NEOs $\Delta t$         & $\mathbf{4620\pm 720}${\bf\,y}   & \textcolor{gray}{$3900\pm 600$\,y}   \\
\end{tabular}
\tablefoot{
Values depend on the assumed size~$D$ of Apophis.
}
\end{table}









\begin{acknowledgements}
M.B. was supported by GACR grant no. 25-16789S of the Czech Science Foundation.
J.D. was supported by GACR grant no. 23-04946S of the Czech Science Foundation.
The work of O.C. was supported by the Czech Science Foundation (grant 25-16507S), the Charles University Research Centre program (No. UNCE/24/SCI/005), and the Ministry of Education, Youth and Sports of the Czech Republic through the e-INFRA CZ (ID:90254).
We thank an anonymous referee for their insightful comments.
\end{acknowledgements}

\bibliographystyle{aa}
\bibliography{references}


\appendix

\section{Comparison to \citet{Farnocchia_2021RNAAS...5..257F}}\label{farnocchia}

If we use the \"Opik theory
and apply it to all NEOs larger than 420\,m,
we obtain the NEO population
$N_{\rm neo} = 3380\pm 300$
\citep{Nesvorny_2024Icar..41716110N},
the probability
$p''' = (44.311\pm 3)\times 10^{-18}\,{\rm km^{-2}\,{\rm y}^{-1}}$,
the flux
$\Phi = (1.5\pm 0.2)\times 10^{-13}\,{\rm km}^{-2}\,{\rm y}^{-1}$,
and the corresponding frequency
once per $(4620\pm 720)\,{\rm y}$.

\begin{figure}[h!]
\centering
\includegraphics[width=7.5cm]{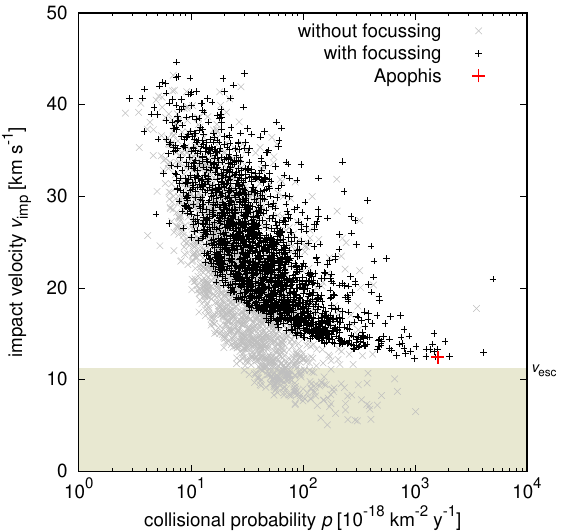}
\caption{
Correlation of the collisional probabilities~$p$
and the impact velocities~$v_{\rm imp}$
computed for NEOs larger than 420\,m.
Our computations were done with ({\bf black})
and without (\textcolor{gray}{gray})
gravitational focussing;
in both cases $p$ and $v$ is negatively correlated.
For reference, the escape speed from Earth,
$11.2\,{\rm km}\,{\rm s}^{-1}$,
is also plotted.
}
\label{correlation_p_v}
\end{figure}

Our value is in modest disagreement with
\citet{Farnocchia_2021RNAAS...5..257F},
who reported once per 7500\,{\rm y}.
The non-negligible difference between the values is due to several factors:
(i)~they used the absolute magnitude limit ($H = 19.1\,{\rm mag}$);
(ii)~their population was 10\% lower ($N_{\rm neo}({<}H) = 3025$);
(iii)~they used the mean, per-object impact probability
$p_1 = 1.66\times 10^{-9}\,{\rm y}^{-1}$,
equivalent to
$p'''' = p_1 / R_\oplus^2 = 40.807\times 10^{-18}\,{\rm km}^{-2}\,{\rm y}$;
(iv)~their gravitational focussing was obtained as an integration over the velocity distribution.
This is different from our, more precise approach,
because we used a per-object impact velocity
\citep{Bottke_1993GeoRL..20..879B},
hence per-object gravitational focussing,
which is correlated with~$p$
(Fig.~\ref{correlation_p_v}).

\section{Verification by an N-body integration}\label{nbody}

If we use a direct, N-body integration
of known NEOs larger than 420\,m,
we obtain a very similar result.
For this particular computation,
we used 3710 orbits from the Astorb catalogue
\cite{Moskovitz_2022A&C....4100661M}.
This number would correspond to the mean geometric albedo
of $p_V = 0.15$
\citep{Mainzer_2011ApJ...743..156M}.
For each orbit, we used 10 clones
with different Yarkovsky drifts
\citep{Vokrouhlicky_1999AJ....118.3049V}
to account for a possibility of encounters.
In this sense, the integration was not deterministic.
Eventually, the number of encounters must be divided by 10.
The overall time span was 10000\,y,
the time step 0.009\,y,
which is sufficient to monitor even the closest encounters with Earth;
we recorded all closer than 0.01\,au.

\begin{figure}[h!]
\centering
\begin{tabular}{ll}
\includegraphics[width=9cm]{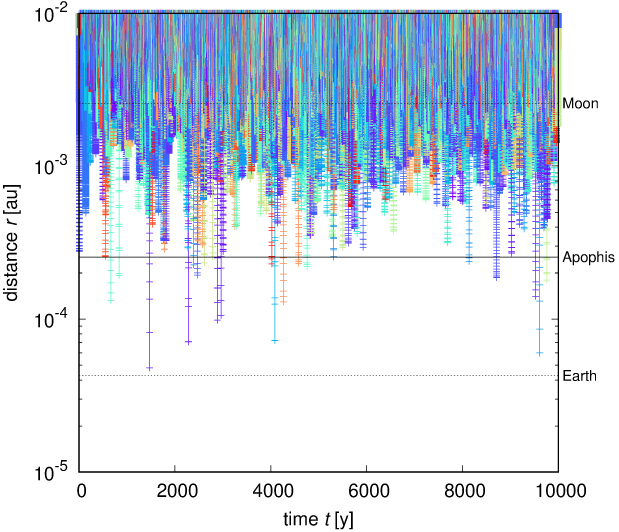} \\
\includegraphics[width=9cm]{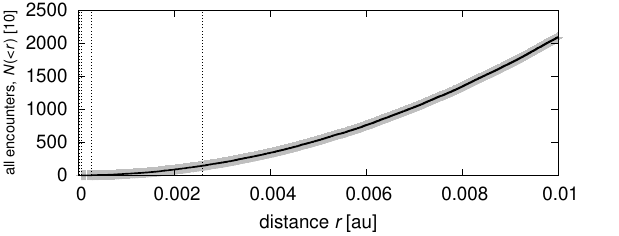} \\
\includegraphics[width=9cm]{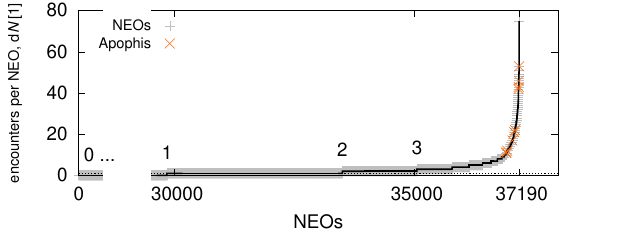} \\
\end{tabular}
\caption{
Top: Close encounter distances vs. time computed for 3710 NEOs
larger than $\sim$420\,m
and their clones (due to the Yarkovsky drift).
The total number of bodies was 37100.
The number of encounters closer than the 2029 Apophis encounter was 21;
the corresponding frequency is approximately
once per 4800\,y.
Middle: Cumulative histogram of distances, $N({<}r)$.
The dependence is parabolic (as $\pi r^2$).
Bottom: Differential histogram of encounters $N_{\rm enc}$ per NEO.
Only 20\%\ of NEOs underwent encounters closer than 0.01\,au
in the course of $10^4\,{\rm y}$.
Specifically, Apophis clones underwent from 11 to 53 encounters
and Apophis-like bodies were responsible for half of all encounters.
}
\label{dahlgren}
\end{figure}

Our result is shown in Fig.~\ref{dahlgren}.
We recorded 20990 encounters,
21/10 = 2.1 of them were closer than the 2029 Apophis encounter
($38000\,{\rm km} \doteq 0.00025\,{\rm au}$).
The corresponding frequency is thus
once per $(4760\pm 1040)\,{\rm y}$,
in agreement with Sec.~\ref{farnocchia}.

\begin{figure}
\centering
\includegraphics[width=8cm]{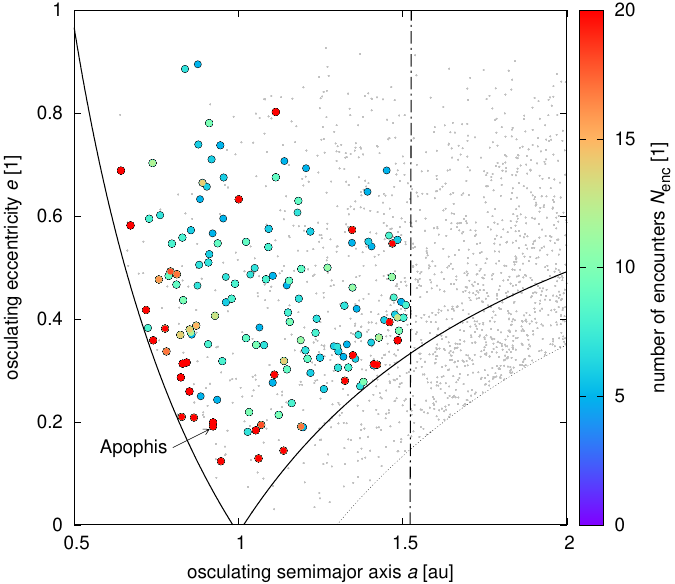}
\caption{
Apophis-like bodies in the NEO space.
Their osculating semimajor axis~$a$ and eccentricity~$e$
fulfill the conditions
$Q > 0.983\,{\rm au}$,
$q < 1.016\,{\rm au}$,
$a < 1.523\,{\rm au}$,
and the number of close encounters with Earth is
$N_{\rm enc}({<}0.01\,{\rm au}) \ge 5$,
computed for 10 clones over $10^4\,{\rm y}$.
Colours correspond to the number of encounters. 
Other NEOs >420\,m are also plotted
(\textcolor{gray}{gray}).
}
\label{apophislike}
\end{figure}


\section{Size from occultations}\label{sec:durech}

The shape model of Apophis used in this work is based
on an extensive set of light curves
\citep{Pravec_2014Icar..233...48P,Durech_2026}.
It is scale-free because visible photometry does
not enable to uniquely determine the albedo.
To scale it, we used occultations of stars by Apophis.
We computed the sky-plane projections of the model
and compared it with the observed occultation timings,
projected on the fundamental plane.
The size was then optimized so that the distance
between the silhouette and the chords was the best
in terms of $\chi^2$.
Details of the method for principal-axis and tumbling rotators
were described in
\citet{Durech_2011Icar..214..652D,Durech_2025A&A...696A..76D}.

\begin{table}
\caption{List of observers, and their locations, who participated in the four occultations shown in Fig.~\ref{durech}.}
\footnotesize
\centering
\begin{tabular}{l}
            \hline
            \multicolumn{1}{c}{7 March 2021} \\[1mm]
            Mark Ziegler, La Salle, CO \\
            Shayne Sivley, La Salle, CO \\
            B. Gowe, Cranbrook, Bc, CA \\
            J. Moore, Oakdale, LA (2 stations) \\
            D.~Dunham \& J. Dunham, Oakdale, LA (5 stations) \\
            R. Venable, Oakdale, LA (4 stations)\\
            Paul Cervantes, Kersey, CO \\
            P. Ceravolo \& D. Ceravolo, Cranbrook, Bc, CA \\
            K. Getrost, Oakdale, LA \\
            Bob McClure, La Salle, CO \\
            Franck Marchis \& Jonathan Horst, La Salle, CO \\
            Ryan Tirashi \& Sharon Tirashi, Deer Trail, CO \\
            Charlie Bicknell \& Christina Bicknell, Strasburg, CO \\
            R. Nugent, Oakdale, LA \\
            N. Carlson, Gage, OK \\
            M. Skrutskie, La Salle, CO \\
            Bob MacArthur, Parker, CO \\
            \\
            \multicolumn{1}{c}{4 April 2021} \\[1mm]
            R. Venable, Chaparral, NM (4 stations) \\
            N. Carlson, San Antonio, NM \\
            K. Getrost, El Paso, TX \\
            \\
            \multicolumn{1}{c}{11 April 2021} \\[1mm]
            N. Carlson, Tucson, NM \\
            V. Sempronio, Sierra Vista, NM \\
            K. Getrost, Clevland, OH \\
            \\
            \multicolumn{1}{c}{6 May 2021} \\[1mm]
            T. Blank, Surprise, AZ (2 stations) \\
            V. Sempronio, Sells, AZ \\
            T. George, Avondale, AZ \\
            N. Carlson, Sells, AZ \\
            S. Aguirre, Hermosillo, Sonora, Mexico \\
            \hline
\end{tabular}
\label{tab:occ}
\end{table}

We used four occultations from 2021
(Tab.~\ref{tab:occ}).
Occultation data were obtained from the PDS Small Bodies Node
\citep{Herald_2024pds..data...77H}.
Because of the small size of Apophis, the duration of occultations
was of the order of 0.1\,s.
Out of the 12 chords,
some have 1-$\sigma$ errors as large as 0.02\,s,
but there are four as small as 0.008\,s,
which constrain the model.

\begin{figure}
\centering
\begin{tabular}{@{}l@{\kern.1cm}l@{}}
\includegraphics[width=4.5cm]{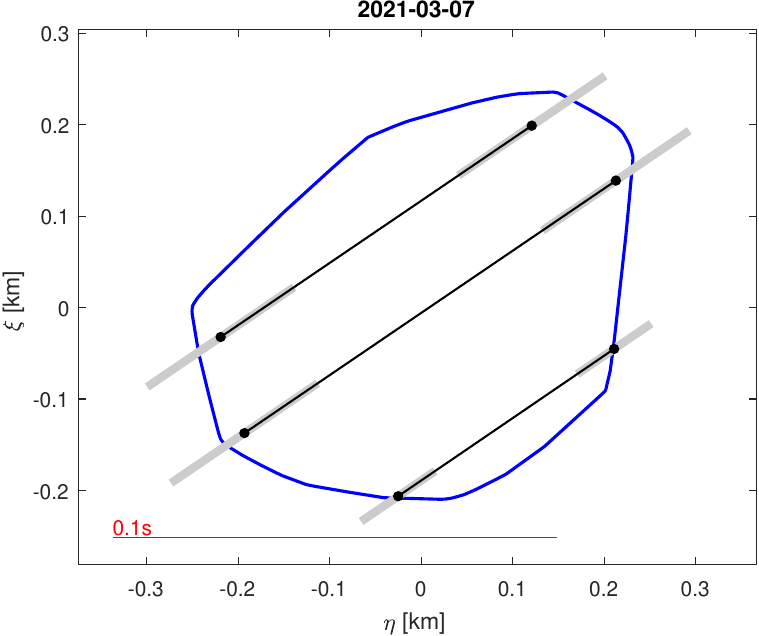} &
\includegraphics[width=4.5cm]{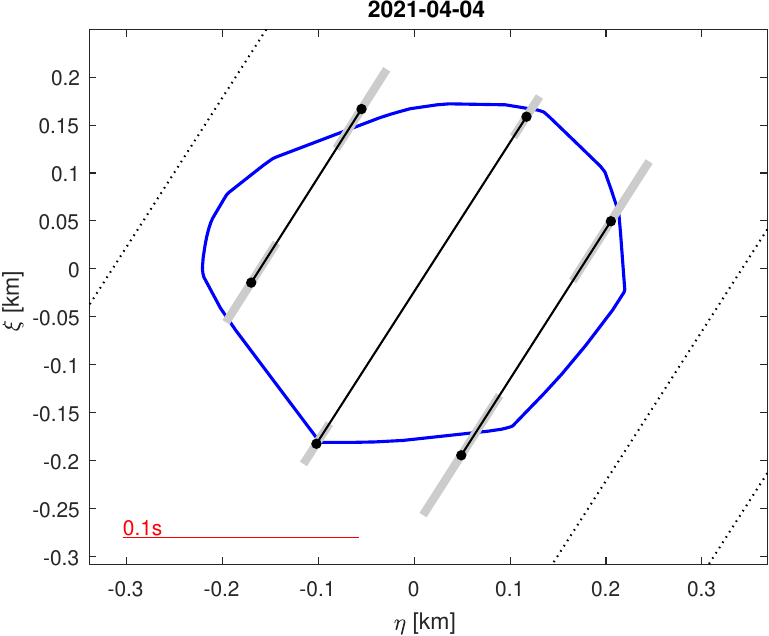} \\
\includegraphics[width=4.5cm]{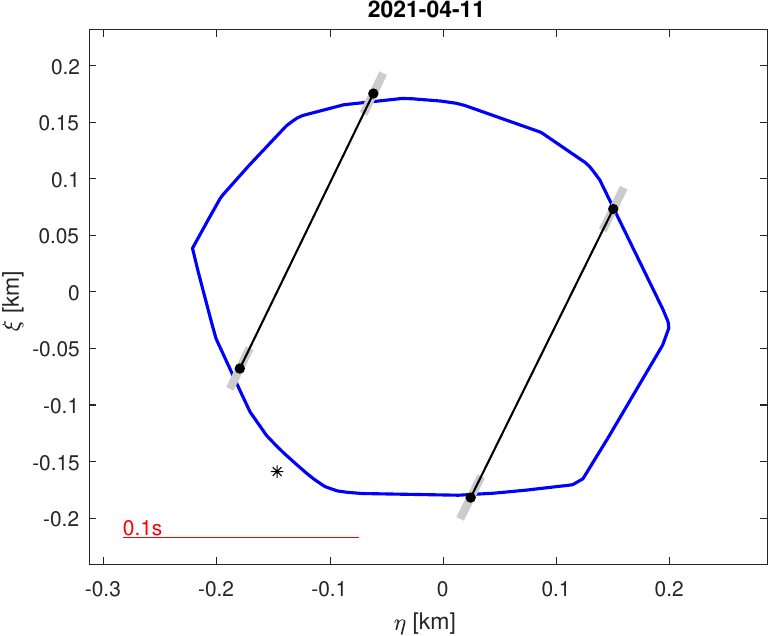} &
\includegraphics[width=4.5cm]{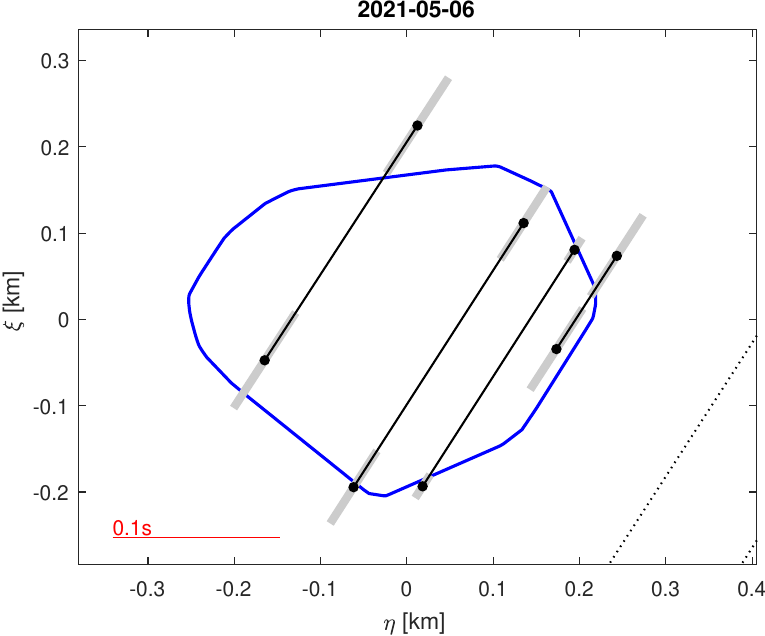} \\
\end{tabular}
\caption{
Convex-hull, light curve shape model of Apophis
(this work),
scaled to the observed occultations.
The resulting volume-equivalent diameter is
$(420\pm 14)\,{\rm m}$.
This model is preferred. 
}
\label{durech}
\end{figure}

Our result is shown in Fig.~\ref{durech}.
The nominal, volume-equivalent size of Apophis is 417\,m,
substantially larger than previous nominal values,
340\,m and 380\,m, respectively
\citep{Brozovic_2018Icar..300..115B,Muller_2014A&A...566A..22M}.
We estimated its uncertainty by a thousand runs
with randomly shifted timings
(according to the reported errors);
the mean value is 420\,m and the standard deviation only 14\,m.

\paragraph{Non-convexity.}
Apart from the timing errors,
there is a contribution from systematic model errors.
In particular, the true shape of Apophis can deviate
from our convex-hull shape.
We checked several known, non-convex shapes
(Eros, Bennu, \dots)
against corresponding convex hulls,
and their difference varied from $0.5$ up to $5\,\%$,
in terms of~$D$.
If true, the size of Apophis decreases down to 400\,m.  

\begin{figure}
\centering
\begin{tabular}{@{}l@{\kern.1cm}l@{}}
\includegraphics[width=4.5cm]{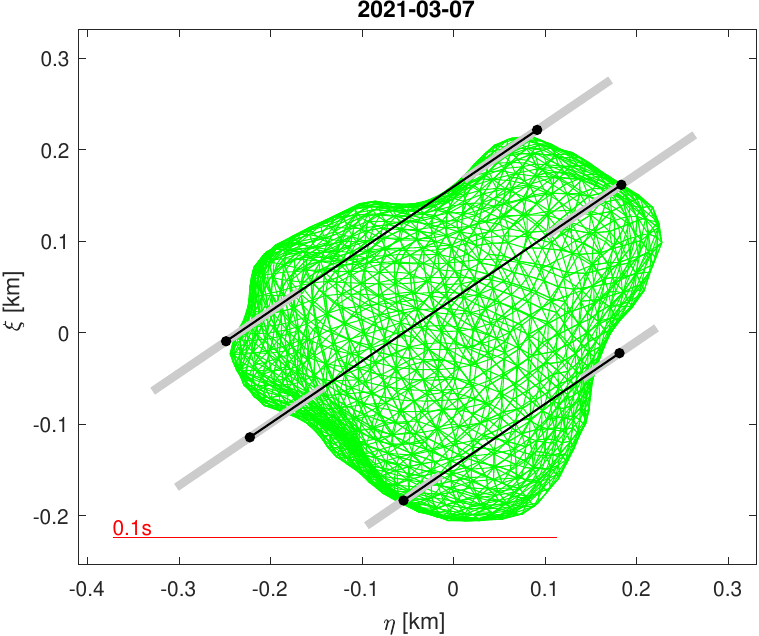} &
\includegraphics[width=4.5cm]{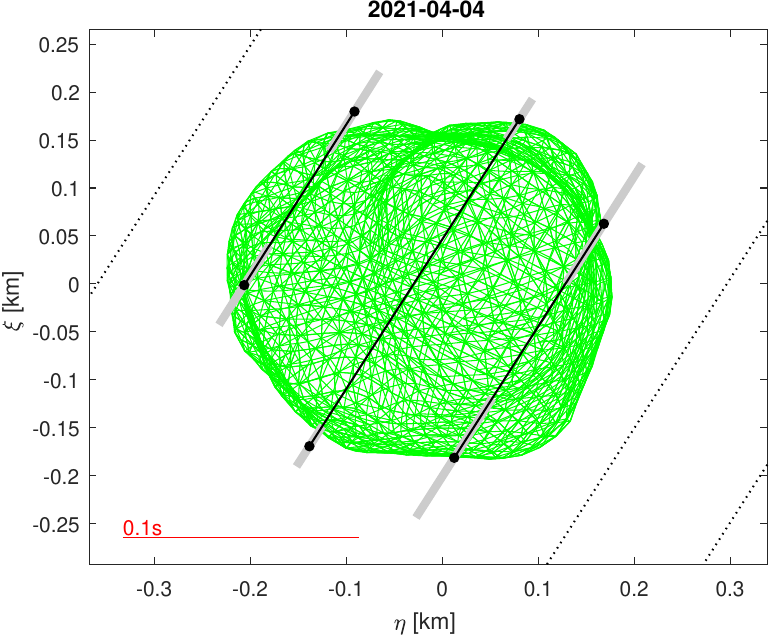} \\
\includegraphics[width=4.5cm]{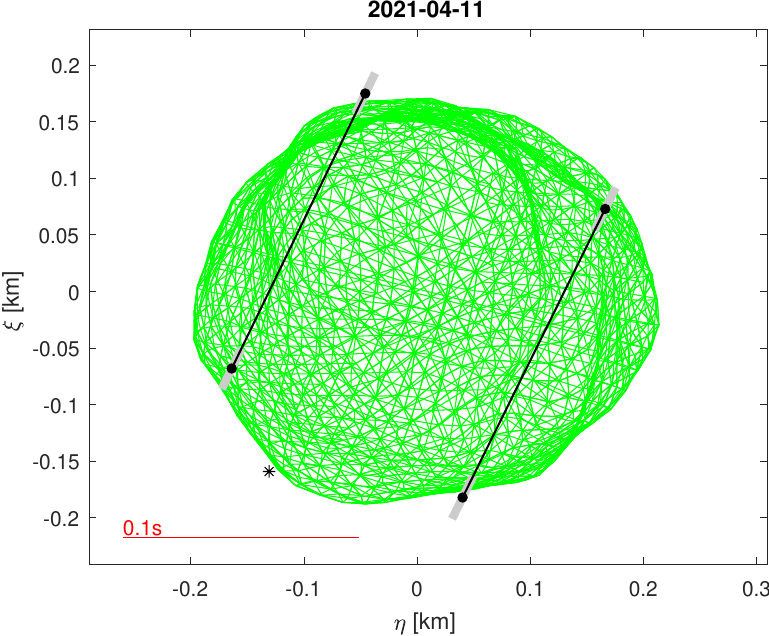} &
\includegraphics[width=4.5cm]{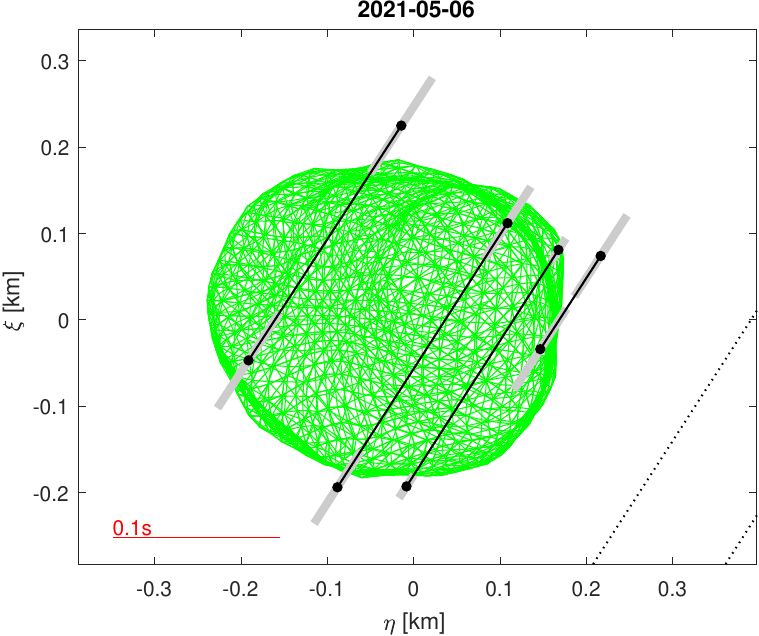} \\
\end{tabular}
\caption{
Non-convex, radar shape model of Apophis \citep{Brozovic_2018Icar..300..115B},
scaled to the observed occultations.
The resulting volume-equivalent diameter is
$(391\pm 14)\,{\rm m}$.
This model is substantially different from Fig.~\ref{durech},
and fits light curves worse than the convex model.
}
\label{durech_radar}
\end{figure}

For comparison,
we scaled the non-convex, radar shape model of Apophis from 
\cite{Brozovic_2018Icar..300..115B}
and its volume-equivalent size is 391\,m
(Fig.~\ref{durech_radar}).
However, it does not fit light curves correctly.
The most drammatic deviation from convexity would be binarity.
Possibly, it is seen as a ``eerie darkness''
in some of the Doppler--delay images
(\citealt{Brozovic_2018Icar..300..115B}, fig.~2).
If true, the size of Apophis decreases further,
down to $\sim$380\,m.

\paragraph{Population estimates.}
The size of Apophis is linked to our population estimates
in Sec.~\ref{sec:mean} and~\ref{farnocchia}.
If it is revised from 420\,m to $\sim$380\,m,
they must be revised as follows.
For the Flora family,
which has a shallow slope ${\sim}{-1.3}$,
$N_{\rm mb} = 23000\pm2000$,
$N_{\rm neo} = 670\pm 150$, and
the frequency is
once per $(12000\pm 3100)\,{\rm y}$.
For all NEOs,
having the slope $-1.9$,
$N_{\rm neo} = 3980\pm 350$, and
the corresponding frequency is
once per $(3900\pm 600)\,{\rm y}$.
None of these revisions would substantially change our conclusions.


\section{Supplementary figures}

\begin{figure}[h!]
\centering
\begin{tabular}{c}
Juno (L/LL) \\
\includegraphics[width=9cm]{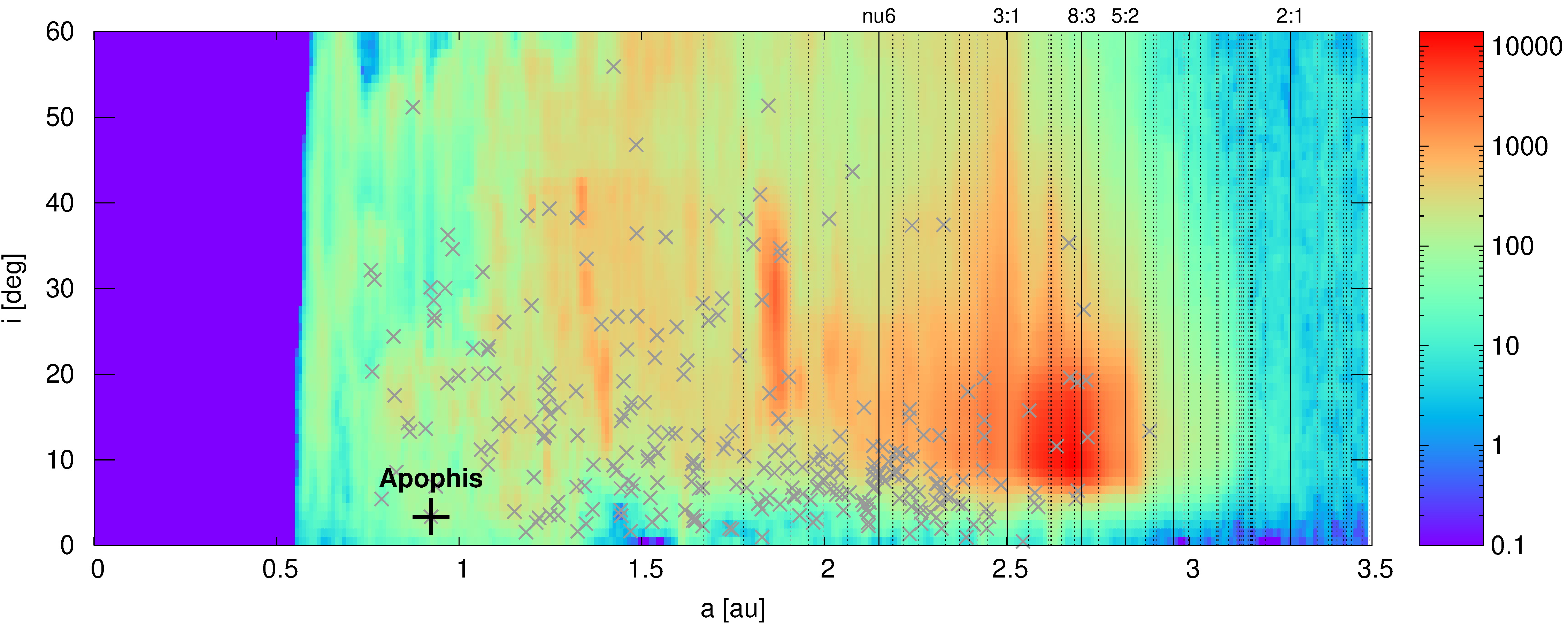} \\
\end{tabular}
\caption{
Same as Fig.~\ref{maps}, but for the Juno family;
its distribution would correspond better
to L/LL- or L-like NEOs.
}
\label{juno}
\end{figure}

\begin{figure}
\begin{tabular}{ll}
\includegraphics[width=8cm]{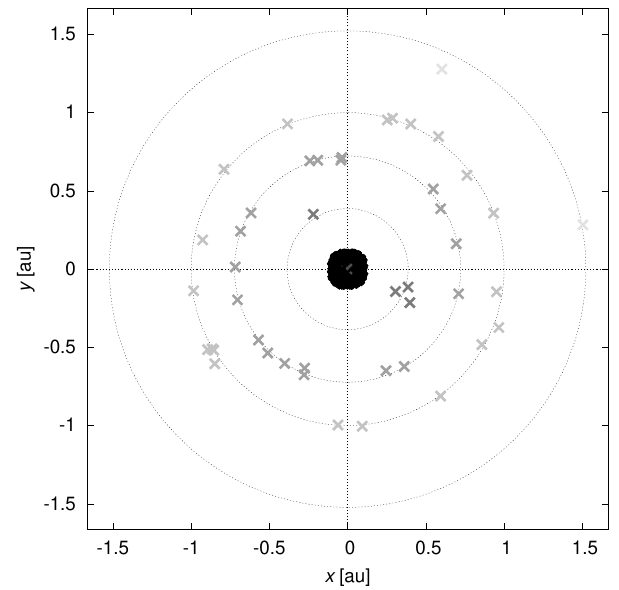} \\
\kern.1cm
\includegraphics[width=7.8cm]{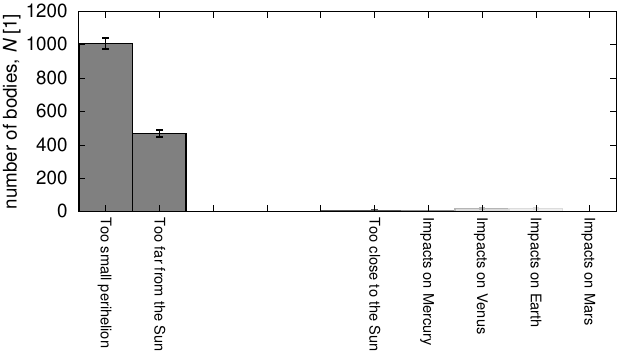} \\
\end{tabular}
\caption{
Same as Fig.~\ref{discard}, but for 3710 NEOs
larger than 420\,m.
In this case, planetary impacts are rare
($3.0\pm 0.4\,\%$).
The time span of simulation (${\sim}5\,{\rm My}$)
corresponded to short-lived NEO orbits.
In equilibrium, they are being replenished
from the belt.
}
\label{nea-3_420m_lclose_0.9d}
\end{figure}

For comparison, we show the Juno family in Fig.~\ref{juno}
and collisions of all NEOs in Fig.~\ref{nea-3_420m_lclose_0.9d}. 

\end{document}